\documentstyle[psfig,twocolumn,eqsecnum,aps]{revtex}

\begin{document}
\draft
\preprint{HEP/123-qed}

\wideabs{
\title {\Large Anomalous Low Temperature States in CeNi$_2$Ge$_2$}
\author {F.~M. Grosche \cite{byline}, P. Agarwal, S.~R. Julian, N.~J. Wilson, R.~K.~W. Haselwimmer, \\ 
S.~J.~S. Lister, N.~D. Mathur, F.~V. Carter, S.~S. Saxena and G.~G. Lonzarich}
\address {Cavendish Laboratory, Madingley Road, Cambridge CB3 0HE, UK}
\date{\today}
\maketitle

\begin{abstract}
\noindent 
Ambient pressure studies on high purity single crystals of the stoichiometric 4f-electron metal CeNi$_2$Ge$_2$ reveal anomalous low temperature forms of the resistivity which challenge our understanding of the metallic state. Comparisons are made with the isostructural and isoelectronic compound CePd$_2$Si$_2$ near the border of magnetism at high pressure, and possible reasons for this novel non-Fermi liquid form of the resistivity are discussed. Phase diagrams of further anomalies are presented, which involve a loss of resistance at low temperature in some samples of CeNi$_2$Ge$_2$ and unexpected high pressure phases.
\end{abstract}

}
\narrowtext
\section {Introduction}
\label{sec:level1}
Strongly correlated electron materials in general and the heavy
fermion compounds in particular exhibit unusual metallic states 
and low temperature phase transitions that remain only partly understood.  
The temperature dependences of the thermodynamic and transport properties,
and in particular of the resistivity, allow us to identify a number of 
apparently distinct regimes.
At high temperatures, a weakly temperature dependent, large resistivity 
consistent with scattering from thermally disordered local magnetic moments 
is observed down to an upper temperature scale, $T_{sf}$ .  Below $T_{sf}$, 
the resistivity drops with decreasing temperature and in the absence of a 
phase transition the resistivity and other bulk properties follow the 
predictions of Fermi liquid theory below a lower temperature scale $T_{FL}$.  
The range between $T_{FL}$ and $T_{sf}$, sometimes described as a spin 
liquid regime, appears as a narrow cross-over region in most materials.

An increasing number of systems are coming to light, however, in which - due 
to their proximity to magnetic phase transitions - the Fermi liquid regime 
is suppressed to very low temperatures or even masked by the onset of 
superconductivity or other forms of order (see e.g. \cite{julian96b}).  
The unconventional normal states observed in these metals might be examined 
in the first instance in terms of phenomenological models for the fluctuations 
of the local order parameter, i.e. the local magnetisation for ferromagnetism 
and the local staggered magnetisation for antiferromagnetism \cite{magninter}, 
or of some associated variable.  If the magnetic ordering temperature is 
suppressed to absolute zero, these modes soften over large portions of reciprocal 
space at low temperatures, leading to a strong enhancement of the quasiparticle 
scattering rate and potentially to a breakdown of the Fermi liquid description 
in its simplest form.  This breakdown may be expected to be particularly 
apparent in the temperature dependence of the resistivity $\rho(T)$ that may 
deviate from the usual Fermi liquid form $\rho(T) \sim T^2$ in pure samples 
at low temperatures.

To look for a breakdown of the Fermi liquid description  in pure materials 
as opposed to the more extensively studied doped \cite{santabarbara96,rosch96} 
heavy fermion systems, we have selected stoichiometric compounds that are 
close to being magnetically ordered at low temperature and have used hydrostatic 
pressure to tune these compounds through quantum ($T \rightarrow 0 K$) phase transitions.

The systems that we have selected, namely the isostructural and isoelectronic 
relatives CePd$_2$Si$_2$ and CeNi$_2$Ge$_2$, allow for examinations of 
an antiferromagnetic quantum critical point in pure metals for the first time
in considerable detail.  
CeNi$_2$Ge$_2$ and CePd$_2$Si$_2$ are isostructural to the heavy fermion 
superconductor CeCu$_2$Si$_2$ \cite{steglich79} and its larger volume relative 
CeCu$_2$Ge$_2$ \cite{jaccard92} (both with the ThCr2Si2 structure), but differ from 
CeCu$_2$Si$_2$ in the number of d electrons in the d-metal constituent, and 
hence in the character of the Fermi surface and in the magnetic properties.

\begin{figure}
\psfig{figure=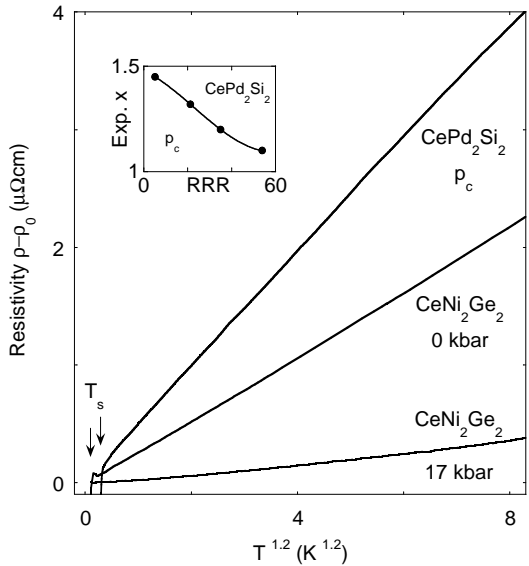,width=3.37in,rheight=3.2in}
\caption{Low temperature resistivity plotted against $T^{1.2}$ (i) 
in CeNi$_2$Ge$_2$ at ambient pressure (denoted everywhere as 0 kbar) 
and at high pressure (17 kbar), and (ii) in CePd$_2$Si$_2$ near 
the critical pressure $p_c \simeq \mbox{28 kbar}$, where 
$T_N \rightarrow 0 K$. For 
comparison, the curves are shifted by their respective $T=0$ intercepts, 
i.e. their residual resistivities, $\rho_0$. For CeNi$_2$Ge$_2$, $\rho_0 \simeq 
0.27\mu\Omega cm$ at $p=0$ and $\simeq 0.23\mu\Omega cm$ at 17 kbar, 
while for CePd$_2$Si$_2$, $\rho_0 \simeq 1.9 \mu\Omega cm$ at 28 kbar. 
Inset: The resistivity exponent $x$, defined by fitting $\rho(T) = \rho_0 + 
A T^x$ from 0.4 K to 4K for a number of samples of CePd$_2$Si$_2$ of 
varying residual resistivity ratio, $\mbox{RRR} = \rho(293K)/\rho_0$. 
Samples with the highest RRR were produced in an RF furnace by quenching
a high purity stoichiometric melt in a water cooled copper boat in UHV
or ultra pure argon atmosphere. The ingots were then annealed at 
approximately 100 $^\circ$C below the melting temperature for one day.
Samples with RRR of over 50 for CePd$_2$Si$_2$ and 300 for CeNi$_2$Ge$_2$ 
were thus obtained.}
\label{normal}
\end{figure}

At ambient pressure, CePd$_2$Si$_2$ orders in an antiferromagnetic
structure with a comparatively small moment of 0.7 $\mu_B$ 
below a N\'{e}el temperature \mbox{$T_N$} of about 10 K \cite
{grier84}, which falls with increasing pressure \cite{thompson86}.
The spin configuration consists of ferromagnetic (110)
planes with spins normal to the planes and alternating in
direction along the spin axis. In a recent study \cite{grosche96},
we have elucidated the phase diagram of CePd$_2$Si$_2$ up to
hydrostatic pressures of about 30 kbar. The N\'eel temperature has
been found to drop linearly with pressure above 15 kbar and to
extrapolate to zero at a critical pressure, $p_c \simeq \mbox{28
kbar}$, while the shoulder in the resistivity, $T_{sf}$, shifts
from 10 K at low pressure to about 100 K near $p_c$.
Superconductivity appears below 430 mK in a limited pressure region
of a few kbar on either side of $p_c$. This behaviour, and perhaps
that in a related system CeRh$_2$Si$_2$  \cite{movshovich96}, is believed to be 
consistent with an anisotropic pairing arising from magnetic
interactions \cite{millis88,mathur98}.
Here, we concentrate on the striking normal state behaviour of the 
resistivity, which deviates strongly from the $T^2$ form usually 
associated with a Fermi-liquid (upper curve in Fig.~\ref{normal} 
and left inset in Fig.~\ref{exponents}). The range between $T_{sf}$ 
and $T_{FL}$, in many materials a narrow cross-over regime, thus
appears to open up to more than two orders of magnitude in
temperature and becomes the dominant feature of the system,
exposing the intervening spin liquid state for closer scrutiny.

The electronically and structurally equivalent compound 
CeNi$_2$Ge$_2$ \cite{cngold}, which is of central interest here, 
has a slightly smaller
lattice constant and its zero pressure behaviour may be 
expected to be similar to that of CePd$_2$Si$_2$ at a pressure 
close to but higher than $p_c$. 

\section {Results}
As shown in Fig.~\ref{normal}, high purity samples of CeNi$_2$Ge$_2$ 
at ambient pressure and of CePd$_2$Si$_2$ at $p_c$ exhibit qualitatively 
similar anomalous temperature dependences of the resistivity.

\begin{figure}
\psfig{figure=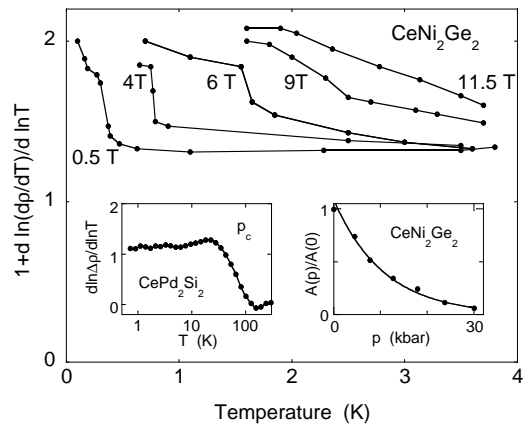,width=3.37in,rheight=2.6in}
\caption{High sensitivity measurement of the 
logarithmic derivative $(1+d \ln (d\rho/dT)/d \ln T)$ of the resistivity
$\rho(T)$ in CeNi$_2$Ge$_2$ at ambient pressure and in applied
magnetic fields. The ordinate reduces to the resistivity exponent $x$
where $\rho(T)$ can be expressed in the form $\rho(T)=\rho_0+A T^x$.
The resistivity was measured along the a-axis with the magnetic field 
applied along the c-axis. The
residual resistivity of the sample was approximately 0.7 $\mu\Omega cm$.
Left inset: The temperature dependence of $d\ln (\Delta\rho = \rho-\rho_0)
/d\ln T$ vs. $T$
for CePd$_2$Si$_2$ close to $p_c$. The ordinate again reduces to the 
resistivity exponent $x$ under the condition given above. In contrast to
$(1+d \ln (d\rho/dT)/d \ln T)$, $d\ln (\Delta\rho)/d\ln T$ is not independent 
of the method of determining the residual resistivity. For reasonable 
estimates of $\rho_0$, however, the two expressions yield the same
result to $\pm 0.1$ over most of the temperature range explored. 
Right inset: Pressure-dependence of the coefficient A, obtained by fitting 
the resistivity of CeNi$_2$Ge$_2$ to the power-law $\rho = \rho_0 + A T^x$ 
from 0.4 K to 10 K.}
\label{exponents}
\end{figure}

For CePd$_2$Si$_2$ near $p_c$, the resistivity has a form 
$\rho = \rho_0 + A T^x$ with an exponent $x$ close to 1 over a wide 
range, from the onset of superconductivity near 0.4 K up to about 
40 K (left inset of Fig.~\ref{exponents}).  Experiments carried out on 
various samples and different sample orientations have revealed a 
variation of the exponents in the range $1.1 < x < 1.4$ and a general 
trend towards lower values for purer (lower $\rho_0$) specimens, indicating 
a possible limiting value of close to 1 for ideally pure samples 
(inset of Fig. 1).

Our present measurements at ambient pressure show that CeNi$_2$Ge$_2$ 
follows a similar power law variation of the resistivity over at least 
an order of magnitude in temperature above about 200 mK 
and up to pressures of at least 17 kbar (Fig.~\ref{normal}). 
We have extended this study to lower temperatures under applied magnetic 
fields in a high quality sample, which showed no sign of a superconducting 
transition down to 100 mK at 0.5 T and above.  At 0.5 T, a detailed analysis 
of the temperature dependence of the power-law exponent (more precisely, 
the logarithmic derivative defined in the caption of Fig.~\ref{exponents}) 
reveals a rapid cross-over to the Fermi-liquid value $x = 2$ below a 
temperature $T_{FL} \simeq \mbox{200 mK}$ (Fig.~\ref{exponents}). This cross-over 
region rises and broadens with increasing magnetic field 
(Fig.~\ref{exponents}). 
These findings indicate that CeNi$_2$Ge$_2$ is 
delicately placed close to the critical point studied in CePd$_2$Si$_2$ 
at high pressure, returning to Fermi liquid behaviour at low temperatures 
as the spin fluctuations are quenched by an increasing magnetic field.

\begin{figure}
\psfig{figure=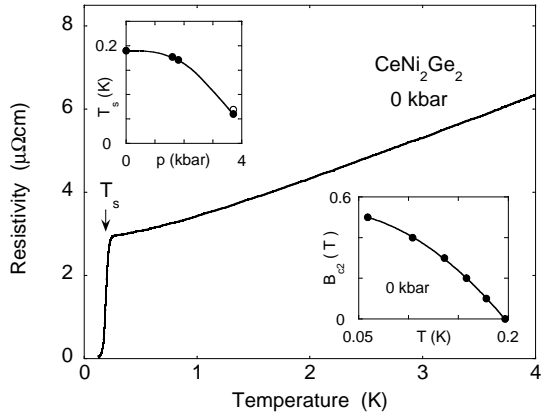,width=3.37in,rheight=2.6in}
\caption{Low temperature resistivity $\rho(T)$ in CeNi$_2$Ge$_2$ at ambient 
pressure in a sample which exhibits a complete loss of resistance below 200 
mK. Upper inset: The variation of the transition temperature $T_s$ with
hydrostatic pressure. $T_s$ is measured at the 50\% drop in $\rho(T)$ 
(solid points), except at the highest pressure where the drop in resistivity
is 14\% (open point) at the lowest temperature reached (the solid point in this case 
was obtained by extrapolation).
No significant drop in resistance was observed above 4 kbar. Lower inset: 
Magnetic field dependence of the transition temperature $T_s$ in another 
sample of CeNi$_2$Ge$_2$, indicating an initial slope 
$dB_{c2}/dT \simeq -5 T/K$.
}
\label{cngsuper}
\end{figure}

One sample of CeNi$_2$Ge$_2$ shows a complete loss of resistance at 
ambient pressure below 200 mK (Fig.~\ref{cngsuper}), similar to the 
occurrence of superconductivity in high pressure CePd$_2$Si$_2$, while 
a number of other high quality crystals exhibit a drop in $\rho(T)$ 
of about 85\% at low temperatures. We note that a downturn of 
approximately 10\% in the resistivity of CeNi$_2$Ge$_2$ below 100 mK 
is also evident from data in \cite{gegenwart98}. A study of the shift 
of this transition in a second sample with magnetic field
yields an initial slope $d B_{c2} /d T \simeq - 5 T/K$ (inset of 
Fig.~\ref{cngsuper}), comparable 
with the value of $\simeq - 6 T/K$ observed for the superconducting 
transition in high-pressure CePd$_2$Si$_2$ \cite{grosche97}.
The transition is very sensitive to hydrostatic pressure 
(inset of Fig.~\ref{cngsuper}) and above 4 kbar, no drop of the 
resistivity is observed. The resulting phase diagram is reminiscent of 
the behaviour of CePd$_2$Si$_2$ at high pressure and is consistent with 
our conjecture that CeNi$_2$Ge$_2$ is a smaller volume relative of 
CePd$_2$Si$_2$.

\begin{figure}
\psfig{figure=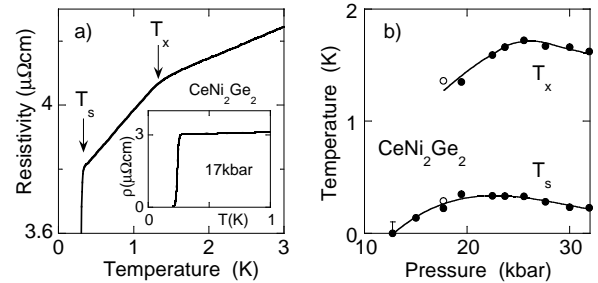,width=3.37in,rheight=1.75in}
\caption{Further high pressure phases in CeNi$_2$Ge$_2$ can be identified from 
anomalies in the form of $\rho(T)$ at low temperature. (a): Low temperature 
resistivity of a sample of CeNi$_2$Ge$_2$. $T_s$ labels a sudden loss in 
resistance, which is complete in a number of samples (inset), while 
$T_x$ refers 
to a kink in the temperature dependence of $\rho(T)$ reminiscent of a magnetic 
phase transition. (b): Pressure dependence of the two anomalies determined
on a second sample (closed symbols) and the data extracted from (a) (open symbols). 
These high pressure
transitions are most pronounced in samples of CeNi$_2$Ge$_2$ produced by slow 
Czochralski pulling from a pure stoichiometric melt in a water cooled Cu boat.
Crystals grown in this way were found to have values of RRR of around 20.
}
\label{cnghi}
\end{figure}

Further anomalies were discovered at still higher pressure in CeNi$_2$Ge$_2$, 
which showed indications of a new ordered phase (labelled $T_x$ in 
Fig.~\ref{cnghi}) at around 1 K and, again, a drop of resistance of up to 
100\% below about 0.4 K ($T_s$ in 
Fig.~\ref{cnghi}), reminiscent of superconductivity.  
We note that early measurements of the specific heat 
in the same region of the phase diagram have revealed an anomalous peak, 
which may be associated with our upper transition at $T_x$ \cite{hellmann96}.  
In contrast to the superconducting phase found in high pressure CePd$_2$Si$_2$ 
and to the corresponding low pressure phase in CeNi$_2$Ge$_2$, these high 
pressure states in CeNi$_2$Ge$_2$ are relatively insensitive to variations 
of lattice density.  In this sense they are reminiscent of the behaviour 
observed in CeCu$_2$Si$_2$, where a superconducting phase persists over a 
wide region of pressure up to about 100 kbar \cite{bellarbi84}.

\section {Discussion}
In both CeNi$_2$Ge$_2$ and CePd$_2$Si$_2$, the temperature dependence of the 
resistivity is characterised over a wide range by a power-law with exponent 
close to 1, and by rapid cross-overs to the high and low temperature forms. This
behaviour is not limited to a critical lattice density alone, but, at least in 
CeNi$_2$Ge$_2$, appears to extend over a considerable range in pressure. 
These properties of our two tetragonal metals contrast sharply with that of the 
cubic antiferromagnet CeIn$_3$ \cite{walker97,mathur98}. In the latter the 
resistivity deviates from the Fermi liquid form only in a very narrow pressure
range near the critical pressure $p_c$ where $T_N \rightarrow \mbox{0 K}$. At $p_c$ and 
in low magnetic fields the resistivity exponent, or more precisely $d \ln (\rho-\rho_0)
/d \ln  T$, grows smoothly with decreasing temperature and tends towards a value of
about $3/2$ at around 1 K.

We now consider to what extent these findings may be understood in terms of the
standard model for spin fluctuation scattering. In this model, the resistivity
is given essentially by the population of spin excitations that is proportional
to an effective volume $q_T^d$ in reciprocal space centred on the ordering wavevector
$\bf Q$. Here $d$ is the spatial dimension and $q_T$ is a characteristic thermal 
wavevector that is proportional to $T^{1/z}$, if the spin fluctuation rate 
$\Gamma_{\bf Q+q}$ at a wavevector ${\bf Q+q}$ varies as $q^z$ at a small $q$ and at 
$p_c$. For the simplest case, $z=2$ and thus the resistivity is predicted
to vary as $T$ for $d=2$ and $T^{3/2}$ for $d=3$ far below $T_{sf}$ and at $p_c$, 
when antiferromagnetic order vanishes continuously \cite{magninter}. These results are 
qualitatively consistent with experiment, if we take $d=3$ for cubic CeIn$_3$,
and if the effective dimension is closer to 2 for tetragonal CePd$_2$Si$_2$
and CeNi$_2$Ge$_2$. The latter assumption is not necessarily inconsistent with the known 
magnetic structure of CePd$_2$Si$_2$, that suggests a frustrated spin coupling along the
c-axis and hence a strongly anisotropic spin fluctuation spectrum \cite{grosche97,mathur98}.

There are at least two major difficulties with this description. Firstly, it assumes that 
essentially all carriers on the Fermis surface scatter strongly from excited spin fluctuations. 
This assumption, however, cannot be justified within the usual Born approximation in the presence 
of harmonic
spin fluctuations in an ideally pure system. Under these conditions, those carriers not
satisfying the `Bragg' condition for scattering from critical spin fluctuations near
$\bf Q$ are only weakly perturbed and at low T lead to a Fermi liquid $T^2$ resistivity and 
{\em not} to the above anomalous exponents \cite{hlubina95}.
Secondly, for the standard form assumed for the temperature and pressure dependence
of $\Gamma_{\bf Q+q}$ the model of the last paragraph predicts (i) a gradual increase
of $d\ln \rho/d\ln T$ with decreasing $T$ tending to the limiting exponent only for 
$T\ll T_{sf}$, and (ii) a rapid cross-over to the Fermi liquid exponent at low T as
a function of pressure (or when the magnetic transition is not continuous). Neither
of these predictions appear to be consistent with our findings in CePd$_2$Si$_2$ 
and CeNi$_2$Ge$_2$.

The first of these two difficulties may perhaps be cured by including the effects of residual 
impurities, spin fluctuation anharmonicity, and corrections to the Born 
approximation that may homogenise the quasiparticle relaxation rate over the
Fermi surface. The second problem might be resolved by means of a more 
realistic model for the temperature and pressure dependences of 
$\Gamma_{\bf Q+q}$ than that currently employed. A first step towards such 
refinements has recently been proposed, specifically for the effects of residual 
impurities \cite{rosch98}, but it is too early to tell whether or not it can
account for all of the features we observe, both in our tetragonal
and the cubic systems, in a consistent way.

We also point out that a description of CePd$_2$Si$_2$ and CeNi$_2$Ge$_2$ 
based on a more extreme separation of charge and spin degrees of
freedom than is present in current approaches, cannot be ruled out 
\cite{coleman98}. A complete theory would have to account not only for the 
differences between our tetragonal and cubic systems and the unexpectedly 
wide range of apparent criticality in CePd$_2$Si$_2$ and CeNi$_2$Ge$_2$, but
also for the occurrence of superconductivity on the border of 
antiferromagnetism in all of these cases, and the higher pressure phases that 
we observe in CeNi$_2$Ge$_2$.

\section{Conclusion}
The 4f-electron metals CeNi$_2$Ge$_2$ and CePd$_2$Si$_2$ offer the possibility 
of observing an unconventional normal state over a wide window in temperature 
and pressure without chemical doping.
The similarity between the two materials suggests that CeNi$_2$Ge$_2$
at ambient pressure is conveniently placed very close to the 
antiferromagnetic quantum critical point, as studied under high 
pressure in CePd$_2$Si$_2$, and make it an attractive material 
for future investigations. 
The fixed value of the power-law exponents over
a wide range in temperature - reminiscent of the
behaviour observed in some of the high T$_c$ oxides - and the wide pressure
range of apparent criticality appear
to defy a description in terms of the spin fluctuation model
in its simplest form. 

Superconductivity in rare-earth based heavy
fermion metals is an uncommon phenomenon. 
CeNi$_2$Ge$_2$ may be the second ambient pressure superconductor in this class 
after CeCu$_2$Si$_2$, if the zero-resistance state observed at low pressures can 
be identified as superconductivity. The pairing mechanism in CePd$_2$Si$_2$ and
CeNi$_2$Ge$_2$, in which superconductivity exists only close to 
the very edge of antiferromagnetic order, may however 
be different from that in CeCu$_2$Si$_2$ \cite{fulde84}.
On the other hand, the high pressure, zero-resistance state in CeNi$_2$Ge$_2$ 
could offer a new perspective for our understanding of the first heavy fermion 
superconductor, CeCu$_2$Si$_2$.

\acknowledgments
We thank, in particular, P. Coleman, J.
Flouquet, P. Gegenwart, C. Geibel, I. Gray, S. Kambe, D. Khmelnitskii, 
F. Kromer, M. Lang, A. P. Mackenzie, G. J.
McMullan, A. J. Millis, P. Monthoux, C. Pfleiderer, A. Rosch, G. Sparn, F. Steglich,
A. Tsvelik and I. R. Walker. The research has been supported partly
by the Cambridge Research Centre in Superconductivity, headed by Y.
Liang, by the EPSRC of the UK, by the EU, and by the Cambridge
Newton Trust.


\end {document}